\documentclass[11pt]{article}
\usepackage[margin=1in]{geometry}
\usepackage{amsmath,amssymb}
\usepackage{graphicx}
\usepackage{booktabs}
\usepackage{hyperref}
\usepackage[british]{babel}
\usepackage{csquotes}
\usepackage{enumitem}
\usepackage[numbers,sort&compress]{natbib}
\usepackage{tikz}

\title{Academics and Generative AI: Empirical and Epistemic Indicators of Policy–Practice Voids}
\author{R. Yamamoto Ravenor \\ Tokyo Women’s Medical University, Japan \\ \texttt{yamamoto.ravenor@twmu.ac.jp}}
\date{} 

\begin{document}

\maketitle

\begin{abstract}
As \emph{generative} AI diffuses through academia, policy–practice divergence becomes consequential, creating demand for auditable indicators of alignment. This study prototypes a ten-item, indirect-elicitation instrument embedded in a structured interpretive framework to surface voids between institutional rules and practitioner AI use. The framework extracts empirical and epistemic signals from academics, yielding three filtered indicators of such voids: (1) \textit{AI-integrated assessment capacity (proxy)} — within a three-signal screen (AI skill, perceived teaching benefit, detection confidence), the share who would fully allow AI in exams; (2) \textit{sector-level necessity (proxy)} — among high output control users who still credit AI with high contribution, the proportion who judge AI capable of challenging established disciplines; and (3) \textit{ontological stance} — among respondents who judge AI different in kind from prior tools, report practice change, and pass a metacognition gate, the split between material and immaterial views as an ontological map aligning procurement claims with evidence classes.

A pilot deployment \(n=214\) illustrates how the instrument yields structured indicators that are interpretable across institutional contexts. Insights from the three indicators are (i) capacity unused in the absence of authorisation; (ii) necessity signals discounted under contested adoption; (iii) procurement approvals arbitrary without rubrics. The indicators are designed for reuse and support assessment pilots, planning, monitoring, staff development, and procurement, offering a minimal, reusable measurement layer.
\end{abstract}

\clearpage


\section{Introduction}

Two operational conundrums frame this paper: policy void of effect, and practice void of authorisation. Divergence between policy and practice is commonplace in academia, but \emph{generative} AI \footnote{Throughout, \enquote*{generative AI} or \enquote*{AI} refers to general-purpose generative tools (e.g., large language models and multimodal generators) used for academic tasks.} renders it operationally consequential. Before generative AI, digital tools sat at the periphery of academic cognition (e.g., search, reference managers, LMS utilities) and could be compartmentalised, whether licensed in labs, restricted with ease, or simply ignored without overt policy conflict. By contrast, generative AI targets core artefacts of higher education (language, computation, analysis), is available off-platform and off-campus, and is embedded in everyday suites, where bans are porous and use is hard to police. This ubiquity strips away the old cover for ambiguity in application, making actual or potential policy–practice voids conspicuous. A mix of national surveys, institutional studies, and qualitative research, largely focused on students, underscore the reality of the phenomenon, but have not distilled the inquiry into a dedicated, compact instrument for measuring it. This study takes the view that academics’ generative AI praxis\footnote{\enquote*{Praxis} here means reflective professional practice; action guided by articulated reasons.} can aid the detection and indirect measurement of such voids. 

\emph{Policy–practice voids} are conditions where policy is void of effect at the point of practice or where practitioner agency is void of authorisation. Concretely, this occurs when policy is silent, ambiguous, contradictory, or unenforced in relation to observed practice; or when practitioners are willing and able to act but lack mandate or access. In such cases, policy and practice fail to jointly determine action. Institutional policy is not observed directly in this study; instead, the indicators quantify prevalence signals of these voids as expressed in practitioner practice. The focus is on \emph{academics}\footnote{Here, understood as in-service or retired educators in higher or tertiary education, including teaching-focused and research-active roles.} because epistemic training, peer-review culture, methodological awareness, and responsibility for assessment give their reflections on AI disproportionate policy relevance.

While a mix of national surveys, institutional reviews, and qualitative research — largely focused on students and relying on direct-response formats — have documented gaps between generative AI policy and practice in academia descriptively, none, as far as is known, have proposed a compact, coded instrument embedded in a structured framework to make such divergence empirically legible and institutionally actionable.

The ten-item, indirect-elicitation survey instrument introduced in this study applies two complementary analytic dimensions: 
\begin{itemize}
  \item \textit{Empirical (practice-focused):} operational signals derived from interaction quality between academics and generative AI tools (Item 1a), degree of output control (1b), perceived contribution (1c), teaching value (Item 2), policy orientation (Item 3), and AI detection confidence (Item 4), as well as a perceived AI-necessity signal (Item 5) filtered through high-control and high-contribution cases.
  \item \textit{Epistemic (concept-focused):} orientation to AI (whether seen as different in kind from prior tools, Item 7, and reported change in practice, Item 6) and to intellect (material vs.\ immaterial, Item 9), with a minimal metacognition criterion (Items 8 or 10) used to determine whether ontological responses are analytically included.
\end{itemize}

Estimates are descriptive, reported as binomial proportions with Wilson 95\% confidence intervals. One planned \(2\times2\) association is summarised by an odds ratio (OR) with Fisher’s exact test, using complete and coherent cases. Each of the three analyses applies filtered prevalence to surface potential policy–practice voids:

\begin{itemize}
  \item \textbf{Analysis 1 (AI-integrated assessment capacity):} To gauge readiness for AI-resilient assessment, this analysis identifies respondents who pass a three-signal screen: they demonstrate AI skill (generating multiple viable outputs), affirm improved teaching with AI, and express confidence in detecting AI. This triad suggests they grasp AI affordances in both student and teacher workflows. Among them, the share who would \emph{fully allow} AI use in exams serves as a proxy for AI-resilient assessment capacity. A high prevalence here may indicate unused capacity in the absence of a mandate.

  \item \textbf{Analysis 2 (sector-level necessity):} This analysis isolates rigorous users — academics who tightly control AI outputs (editing, rewriting) yet still credit AI with high contribution. Beyond critical engagement and selective trust, their practices suggest AI is substantially useful. Among them, the share who believe AI could challenge the relevance of established disciplines probes whether AI is seen as \textit{more than useful} — a proxy for perceived sector-level necessity. In contested adoption environments, such necessity signals may be overlooked or discounted.

  \item \textbf{Analysis 3 (ontological stance):} This analysis focuses on respondents who (1) view AI as \emph{different in kind from prior tools}, and (2) report AI-induced change in their academic practice, suggesting reflective engagement and direct experiential impact. Those who also pass a metacognition gate (express prompt-sharing willingness or a brief reflection) are included in a final split: \emph{material} vs.\ \emph{immaterial} views of intellect. This ontological stance constrains which AI procurement claims are epistemically coherent. From a material standpoint, the human intellect is an emergent property of matter organised in a particular way. This view grants that the difference between intellect and AI is, at least in principle, a matter of degree (quantitative), and thus best supports instrumental procurement claims (e.g., throughput, accuracy, scale). An immaterial view holds that AI is \emph{different in kind from the human intellect}, permitting formation-oriented claims (e.g., judgment, synthesis, or disciplinary formation), requiring different evidentiary standards. A simple rubric, crossing purpose (exploration/qualitative vs.\ scale/quantitative) with claim strength (indispensable/need vs.\ beneficial/want), is provided in the \emph{Discussion} section to match coherent claims to evidence classes while avoiding metaphysical adjudication. Without rubrics that connect such claims to appropriate evidentiary standards, procurement risks becoming inconsistent or misaligned.
\end{itemize} 

The aim of this study is methodological: to prototype a minimal, auditable measurement layer that can reveal policy-practice divergence in the academic uptake of generative AI. The ten-item instrument is part of a broader interpretive framework: a set of filters, logics, and rubrics designed to extract, structure, and interpret signals from respondent practice and belief. The present survey serves as an illustrative case, showing item behaviour, pre-specified coding, and estimation logic without claiming population representativeness or causal inference.

The contributions are: 
(i) a compact, ten-item instrument embedded in a structured interpretive framework, designed to surface policy–practice divergence in generative AI academic uptake; 
(ii) a method for generating reproducible indicators that institutions can use to support assessment planning, monitoring, and staff development; and 
(iii) an illustrative deployment showing how filtered response logic can yield interpretable signals for decision-making, including procurement claims aligned to coherent evidentiary standards.

\section{Related Work}

Classic scholarship on implementation and organisation explains why, in academia, written policy may fail to determine practice: universities are \emph{loosely coupled} systems, with autonomy distributed across levels and units \cite{weick1976educational}, and frontline practitioners exercise discretion \cite{lipsky1980street}. Contemporary syntheses reaffirm that policy texts are interpreted and operationalised locally rather than applied verbatim \cite{sin2014policy,trowler2014higher} and update these dynamics for current governance contexts \cite{hautala2018loose,hupe2015introduction}. However, in the context of generative AI, alignment between policy and classroom practice becomes especially critical for assessment integrity and governance \cite{lye2024generative}. UNESCO’s global guidance urges clear principles, safeguards, and institutional capacity-building for human-centred use of generative AI \cite{holmes2023guidance}. Sector guidance (e.g., the Russell Group’s principles; Jisc’s assessment advice) likewise emphasizes preparedness and assessment redesign over prohibition \cite{russell2023russell,jisc2023principles}. Regulators caution that detection-led enforcement is insufficient and encourage authentic, secure assessment approaches \cite{teqsa2025gen}. High student AI uptake intensifies the need for clear institutional positions; UK survey data indicate use for coursework rose from 53\% to 88\% in a year \cite{freeman2025student}. Cross-system examples (e.g., the University of Tokyo; University of Tsukuba) illustrate course-level policies that devolve decisions to instructors, centring on practitioner judgment \cite{utokyo2023gen,tsukuba2024gen}. Broader policy reviews (OECD) similarly stress bridging policy-to-practice processes \cite{schlicht2024bridging,figueroa2024education}.

Empirical work with instructors documents uneven experimentation, governance ambiguity, and calls for support. A multi-institution interview study (Ithaka S+R) reports active exploration alongside demands for policy clarity, AI literacy, and discipline-specific support \cite{baytas2025making}. Other qualitative and mixed-methods studies similarly describe optimism tempered by concerns about integrity and equity, with frequent requests for clear rules and professional development \cite{verboom2025perceptions,sah2025generative,almisad2025faculty,kim2025academic}. At the institutional level, recent scholarship reviews global adoption strategies \cite{jin2025generative} and proposes assessment frameworks attuned to generative AI \cite{ilieva2025framework}.

At a more conceptual level, this study intersects with work that classifies AI use in education. Despite abundant commentary on students and general-purpose typologies, educator-specific classifications of users based on generative-AI practice remain sparse. Examples include conceptual lenses that propose broad archetypes of generative-AI use across domains \cite{riemer2024conceptualizing}, and an obliquely related thread that categorises teacher-users of facial emotion recognition AI tools by orientation, condition, and preference \cite{yamamotoravenor2024ai}, illustrating how user typologies can be constructed in instructional contexts. While adjacent literature often connects measured dispositions to resourcing \cite{venkatesh2003user}, no prior work was found that uses ontological views of the intellect to guide AI provisioning decisions.

A final strand concerns how scholars have understood \emph{intellect} itself (relevant to Item~9 in the survey). In classical philosophy, Platonist views treat \enquote*{intelligibles} (i.e. forms or ideas) as independent of the mind, whereas for Aristotelians the \enquote*{intelligibles} exist \textit{in} the mind, as acts or habits of understanding abstracted from experience \cite{plato1993phaedo,aristotle2017deanima}. Medieval and early-modern debates reprise these contrasts: Thomistic theological accounts argue for immaterial intellectual acts \cite{aquinas1964summa}, Cartesian dualism posits distinct mental and material substances \cite{descartes1996meditations}, while materialist/mechanistic strands deny any immaterial power \cite{lamettrie1996machine}. These orientations endure, often implicitly: in higher education, especially with generative AI, they shape what practitioners take to count as understanding, what they expect AI to contribute, and how they frame the relation between human and machine cognition in teaching and research. In psychological and cognitive traditions, \emph{intellect} is typically operationalised through measurable capacities (e.g., reasoning, concept learning), yielding predominantly materialist models; yet there remain currents in philosophical psychology that defend the immaterial view of the intellect \cite{adler1990intellect}. Separately, contemporary political-economy accounts use \enquote*{immaterial} to characterise knowledge production and \emph{the common} (non-rival, shareable value) in education \cite{lazzarato1996immaterial,lewis2012exopedagogy}, noted here only to mark terminological breadth. Survey Item~9 intentionally does not constrain respondents’ readings of \enquote*{material/immaterial}; it functions as a coarse self-placement that different traditions can underwrite. Regardless of the respondent’s reading of the words \emph{material/immaterial}, the indicator is interpreted strictly as signalling a view about AI–human relation: \emph{difference in degree} (material) versus \emph{difference in kind} (immaterial).

In short, while many studies document generative AI uptake descriptively and encourage institutional alignment, no prior work has introduced a compact, auditable instrument capable of surfacing policy–practice voids through indirect-elicitation indicators. Nor has a typology been proposed that jointly captures operational capacity and epistemic orientation among academics. The framework and indicator set introduced here address both.

\section{Methods} 

\subsection{Conceptual frame}
Let $P$ denote the presence of a clear institutional policy or mandate, $A$ the application and enforcement at the point of practice, and $R$ practitioner capacity (defined as willingness \emph{and} capability to act). Policy is considered “in effect at practice” if and only if all three conditions hold simultaneously:
\[
\text{Policy in effect at practice} \;\iff\; P \land A \land R .
\]
\noindent\hspace*{\parindent}Breakdowns of this condition are labeled as $V_1, V_2,$ etc., where each $V_i$ denotes a practitioner-visible type of policy-practice void. For example:
\[
\begin{aligned}
V_1 &: P \land \neg A &&\text{(policy without effect)},\\
V_2 &: R \land \neg P &&\text{(practice without authorisation)} .
\end{aligned}
\]
\noindent\hspace*{\parindent}Further instances arise from policy silence, ambiguity, or delay ($\neg P$), or from cases where practitioner signals are present but not recognised or acted on. Since neither $P$ nor $A$ is observed directly, the analysis focuses on estimating practitioner-side prevalences related to $R$, alongside epistemic profiles that help surface potential voids.

\subsection{Design and instrument}
This was a cross-sectional, web-based, indirect-elicitation survey of academics who use generative AI in their work ($n=214$). The ten-item instrument was primarily structured around practical questions or prompts. While some items took the form of questions, others elicited numerical estimates, judgments, or brief written responses. Each item was mapped \emph{a priori} to either the empirical (practice-focused) or epistemic (concept-focused) analytic dimensions introduced earlier.

Direct items such as “Are you ready to fully allow AI?” or “Is AI necessary?” are susceptible to social desirability bias, acquiescence, and cautious signalling, especially where institutional policy is unsettled. To reduce these effects, the instrument uses indirect elicitation: separate, grounded items whose joint patterns reveal behaviourally meaningful signals — e.g., AI skill, perceived teaching value, confidence in detection, control-adjusted contribution estimates, and willingness to share prompts or reflect in writing. These signals reduce demand characteristics, enable internal coherence checks across heterogeneous response types, and yield transparent, reproducible prevalence indicators via simple binomial prevalence estimates with confidence intervals. 

Item~9 is intentionally broad; for analysis it is interpreted solely as indicating a \emph{material} (difference-in-degree) or \emph{immaterial} (difference‐in‐kind) view of the relation between AI and the human intellect. No attempt is made to adjudicate between philosophical, psychological, or political–economic traditions.

The survey was web-based (single-page HTML form), bilingual (English and Japanese), and open for participation from December 2024 to October 2025. It was distributed via direct e-mail and academic networks, with a request to forward — constituting non-probability convenience and snowball sampling. No incentives were offered. Eligibility was self-attested: respondents had to be university teachers who use or have used generative AI for work.

Items were drafted in English and rendered into Japanese by a bilingual academic translator. Formatting and response options were aligned across languages. The English wording appears in Appendix~A; the Japanese version is available upon request.

An institutional determination obtained prior to data collection indicated that no formal ethics review was required for anonymous, non-interventional survey research. Participation was voluntary, and analysis used only de-identified records. Submissions were stored on an externally hosted SQL server; routine logs were not linked to survey entries. Optional contact data (if provided) were excluded from the analysis.

Item types were heterogeneous: numeric response (Item~1a), percentage (Item~1c), ordinal choice (Item~1b: Never $\to$ Always), nominal choice (Items 2--9), and open-ended text (Item~10). Analyses used simple summaries matched to item type: binomial proportions (with Wilson 95\% CIs) for binary indicators; conditional proportions for multi-category subgroups; and Fisher’s exact test for a single planned $2\times 2$ association. Item~10 served only as a supplementary, minimal length gate for metacognition and was not subject to content analysis.

Only Item~10 was optional. Missing and coherence rules are detailed in subsection~\emph{Missingness and coherence} below. No de-duplication beyond identical record checks was performed.

\subsection{Operational definitions (pre-specified)}
Common recoding\footnote{Some response labels are abbreviated in text.}: Item~1a (0--10) $\Rightarrow$ \emph{AI-skilled} if $>5$; Item~1b: \emph{high control} is defined as responses of Always or Frequently; Item~1c (0--100) $\Rightarrow$ \emph{high AI contribution} if $\ge 66$ (and set missing if Item~1b = Never); Item~2 and Item~4: Yes/No/Unsure; Item~3: Exam policy preference (Fully allow/Limit/Forbid/Unsure), where \emph{Fully allow} ($3_i = \text{Fully allow}$) indicates no restrictions on AI use; Item~5: “AI could challenge relevance of established disciplines” (Yes/No/Unsure); Item~6: reported \emph{change} in academic practice (Yes/No/Unsure); Item~7: \emph{different in kind} (Only degree/In kind/Unsure); Item~8: prompt-sharing willingness (Yes/Need more information/No); Item~9: view of intellect (Material/Immaterial/Unsure); Item~10: free text; \emph{metacognitive articulation} is defined as a free-text comment of $\ge 20$ characters after trimming whitespace.

\emph{Analysis 1:} AI-integrated assessment capacity (proxy). Respondents who show AI skill, perceive teaching benefit, and report detection confidence — are identified using the indicator variable $T_i$:
\[
T_i = \mathbf{1}\{1\text{a}_i > 5 \land 2_i = \text{Yes} \land 4_i = \text{Yes}\}
\]
\noindent
\noindent\hspace*{\parindent}Here, $\mathbf{1}\{\cdot\}$ denotes the indicator function, which returns 1 if the condition inside is true, and 0 otherwise. The share who would fully allow AI in exams ($Y^{(1)}_i$) is computed as:
\[
Y^{(1)}_i=\mathbf{1}\{3_i=\text{Fully allow}\},\qquad
\]
\[
\hat{p}_{1}=\frac{\sum_i T_i\, Y^{(1)}_i}{\sum_i T_i}.
\]
\noindent\hspace*{\parindent}Here, $\hat{p}_1$ estimates the proportion of respondents meeting all three conditions who would fully allow student use of AI in exams.

\emph{Analysis 2:} Sector-level necessity (proxy). Let $H_i$ identify respondents with high output control, and $C_i$ those who still credit AI with high contribution:
\[
H_i=\mathbf{1}\{1\text{b}_i\in\{\text{Always, Frequently}\}\},\qquad
\]
\[
C_i=\mathbf{1}\{1\text{c}_i\ge 66\},\qquad
\]
\[
\hat{p}_{2a}=\frac{\sum_i H_i\, C_i}{\sum_i H_i}.
\]
\noindent
\noindent\hspace*{\parindent}$\hat{p}_{2a}$ thus estimates the proportion of high-control respondents who still credit AI with high contribution. Among this group ($H_i = 1$ and $C_i = 1$), $N_i$ captures whether they believe AI could challenge established disciplines:
\[
N_i=\mathbf{1}\{5_i=\text{Yes}\},\qquad
\]
\[
\hat{p}_{2b}=\frac{\sum_i H_i\, C_i\, N_i}{\sum_i H_i\, C_i}.
\]

\emph{Analysis 3:} Ontological stance. Respondents who both report practice change and view AI as different in kind are flagged by $E_i$:
\[
E_i = \mathbf{1}\{7_i = \text{In kind} \land 6_i = \text{Yes}\}
\]
\noindent\hspace*{\parindent}Define the metacognition gate, $M_i$, for respondents willing to share prompts or who provided a free-text response of at least 20 characters (denoted $L_i$):
\[
L_i = \text{character count of Item~10 (free-text), after trimming whitespace}
\]
\[
M_i = \mathbf{1}\{8_i = \text{Yes} \lor L_i \ge 20\}
\]
\[
\hat{p}_{3a} = \frac{\sum_i E_i\, M_i}{\sum_i E_i}
\]

Among these, the distribution of Item~9 responses (Material / Immaterial / Unsure) is reported. A planned association between metacognitive articulation and immaterial stance was tested using Fisher’s exact test and odds ratios.

\smallskip
\emph{Note.} Analysis~3 spans two interpretative registers: Item~7 concerns how AI differs from prior computational tools (a technological distinction: kind vs.\ degree), whereas Item~9 concerns AI in relation to the human intellect (a philosophical distinction: kind vs.\ degree).

\subsection{Missingness and coherence}
Analyses were conducted on complete and internally coherent cases for the items required by each indicator. Structural skip: if Item~1b = Never, then Item~1c was treated as inapplicable and excluded. The form constrained Item~1c input to values between 0--100; no invalid entries were observed. Free-text length for Item~10 was computed after trimming whitespace. \enquote*{Unsure} responses were treated as substantive and retained as non-affirmative categories. These pre-specified screens mitigate obvious contradictions and leverage heterogeneous item types (numeric, ordinal, nominal, and text).

\subsection{Estimators and uncertainty}
All point estimates are simple proportions, estimated as $\hat{p} = k/n$, with Wilson 95\% confidence intervals. For the planned $2\times2$ association in Analysis~3 (metacognition $\times$  immaterial stance) Fisher’s exact test (two-sided) along with an exact odds-ratio interval. This association was descriptive only and did not affect other estimators or analytic steps.

\subsection{Scope and reproducibility}
All thresholds and inclusion criteria were specified in advance for clarity and reproducibility. Replication requires only the recoding definitions and the expressions for $\hat{p}_1$, $\hat{p}_{2a}$/$\hat{p}_{2b}$, and $\hat{p}_{3a}$, as detailed earlier. Analyses were implemented in Python using \texttt{pandas}, \texttt{NumPy}, \texttt{SciPy}, and \texttt{statsmodels}.

\section{Results}

These are illustrative results from a pilot deployment. All estimates are descriptive proportions with two-sided Wilson 95\% confidence intervals (CIs), shown here to demonstrate indicator logic and behaviour, not to support population inference. Denominators reflect the pre-specified gates for each analysis; numerators are the subset meeting the outcome.

\paragraph{Analysis 1 — \textit{AI-integrated assessment capacity (proxy)}}
Among respondents demonstrating AI skill (Item~1a $>5$), perceived teaching benefit (Item 2 = Yes), and confidence in distinguishing AI from humans (Item 4 = Yes), 32 out of 58 (55\%) would fully allow AI use in exams (95\% CI: 43–67).

\paragraph{Analysis 2 — \textit{Sector-level necessity (proxy)}}
Among high-control users (Item 1b = Always/Frequently) with valid contribution estimates, 43 out of 195 (22\%) credited AI with a high contribution ($\ge 66$\%; 95\% CI: 17–28). Of those, 35 out of 43 (81\%) agreed that AI could challenge established disciplines (95\% CI: 67–90).

\paragraph{Analysis 3 — \textit{Ontological stance (descriptive indicator)}}
In the ontological indicator set (those who both perceived AI as different in kind from prior tools and reported practice change; $n=119$), 44 respondents (37\%; 95\% CI: 29–46) passed the metacognition gate (either expressed prompt-sharing willingness or wrote a comment of $\ge$ 20 characters). Among these, 41 out of 44 (93\%) selected the immaterial view of intellect (95\% CI: 82–98). A $2\times 2$ table shows a strong association between metacognition and immaterial stance (OR $\approx$ 33; Fisher’s exact $p<0.0001$).

\begin{table}[h!]
\centering
\caption{Summary of key prevalence estimates with Wilson 95\% CIs}
\begin{tabular}{lccc}
\toprule
Panel & $k$ & $n$ & \% (95\% CI) \\
\midrule
A1 — AI-integrated assessment capacity (Fully allow) & 32 & 58 & 55 (43--67) \\
A2a — High control → High AI contribution ($\ge 66\%$) & 43 & 195 & 22 (17--28) \\
A2b — Among A2a: AI could challenge disciplines & 35 & 43 & 81 (67--90) \\
A3a — Ontological indicator set: Metacognition present & 44 & 119 & 37 (29--46) \\
A3b — Among A3a: Immaterial view of intellect & 41 & 44 & 93 (82--98) \\
\bottomrule
\end{tabular}
\end{table}

\begin{table}[h!]
\centering
\caption{Analysis 3 cross-tab: Metacognition $\times$ Immaterial ($n=119$)}
\begin{tabular}{lcc}
\toprule
 & Immaterial & Not immaterial \\
\midrule
Metacognition present ($M{=}1$) & 41 & 3 \\
Metacognition absent ($M{=}0$) & 22 & 53 \\
\midrule
$\text{OR} \approx 33,\quad p < 0.0001\ \text{(Fisher's exact test)}$ & & \\
\bottomrule
\end{tabular}
\end{table}

\noindent These estimates complete the three pre-specified indicators; no additional planned analyses were conducted.

\section{Discussion}

This study piloted a compact, indirect-elicitation instrument to generate structured indicators of generative AI use and epistemic stance among academics. While the survey was illustrative rather than representative, the indicators it yielded reveal practitioner-side capacities and dispositions with clear institutional implications for assessment, curriculum, and procurement. 

\subsection{Implications for institutions}

\begin{itemize}
  \item \textit{Assessment pilots (A1).} Departments with a non-trivial A1 share may unlock AI-integrated assessment capacity by working with staff already willing and able to proceed, even where institutional rules have not yet adapted.

  \item \textit{Curriculum reflection (A2).} A2 signals the presence of academics who see generative AI as increasingly necessary for education to remain relevant. Institutions need not agree, but should at minimum be aware that such a group exists, and eventually make explicit whether current rules treat them as admissible or marginal.

  \item \textit{Procurement review (A3).} A3 helps distinguish which kinds of AI procurement claims are internally coherent, reducing interpretation drift. Figure~\ref{fig:claim-space} summarises this distinction visually, mapping procurement claims by purpose and strength.

\hspace{1cm}
  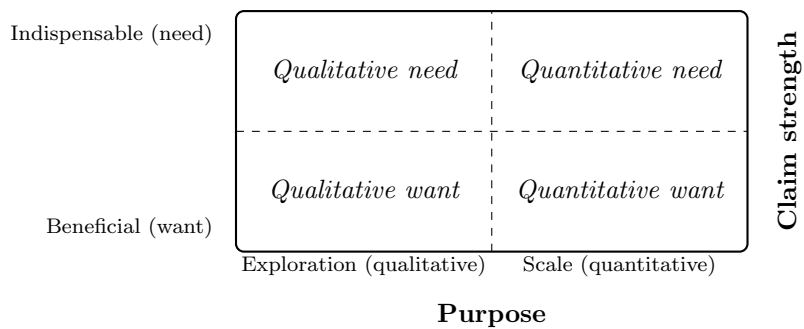
\begin{figure}[ht]
  \centering
  \hspace{0.8cm}
  \begin{tikzpicture}[x=3.4cm,y=1.6cm]
    \draw[thick,rounded corners=3pt] (0,0) rectangle (2,2);
    \draw[dashed] (1,0) -- (1,2);
    \draw[dashed] (0,1) -- (2,1);
    \node[anchor=east] at (-0.05,1.8) {\scriptsize Indispensable (need)};
    \node[anchor=east] at (-0.05,0.2) {\scriptsize Beneficial (want)};
    \node at (0.5,-0.12) {\scriptsize Exploration (qualitative)};
    \node at (1.5,-0.12) {\scriptsize Scale (quantitative)};
    \node[rotate=90, anchor=south] at (2.25,1) {\small\bfseries Claim strength};
    \node[anchor=north] at (1,-0.35) {\small\bfseries Purpose};
    \node[align=center, font=\small\itshape] at (0.5,1.5) {Qualitative need};
    \node[align=center, font=\small\itshape] at (1.5,1.5) {Quantitative need};
    \node[align=center, font=\small\itshape] at (0.5,0.5) {Qualitative want};
    \node[align=center, font=\small\itshape] at (1.5,0.5) {Quantitative want};
  \end{tikzpicture}
  \caption{Space of procurement claims by purpose and strength. The A3 indicator constrains which quadrants are coherent for different ontological stances.}
  \label{fig:claim-space}
  \end{figure}

  When AI procurement claims are made, ontological stance constrains the kinds of reasons that are coherent. A \enquote*{materialist} can plausibly frame a request as an objective \emph{need} only when pointing to quantitative gains. Otherwise, the claim risks being read as a subjective preference, \enquote*{a want}. An \enquote*{immaterialist} can plausibly frame needs in either quantitative or qualitative terms. For interpretive clarity, “indispensable” is mapped to claims that plausibly express \emph{needs}, and “beneficial” to those that plausibly express \emph{wants}. Likewise, “exploration” refers to qualitative aims (e.g., formation, novelty), and “scale” to quantitative aims (e.g., throughput, complexity).

  \item \textit{Monitoring.} Because the indicators are transparent proportions, they support longitudinal tracking and cohort comparison. Over time, publishing the rubric makes drift easier to detect \cite{campbell1979assessing,goodhart1984problems}.
\end{itemize}

\subsection{Strengths and limitations}

This study introduces a deliberately minimal, auditable measurement layer to quantify signals relevant to policy–practice voids in the academic uptake of generative AI. Its strengths are conceptual and methodological. First, the instrument operationalises indirect-elicitation to mitigate social desirability and ambiguous signalling. Second, coding rules were pre-specified, coherence was enforced across heterogeneous item types, and analyses remained descriptive, with thresholds chosen for clarity. Third, the indicators work as modular building blocks, compact enough for reuse, yet interpretable at the level of institutional planning and procurement.

Limitations follow directly from the study’s scope. The sample was non-probability and forward-distributed, limiting generalisability and precluding a response-rate estimate. The thresholds (e.g., Item~1a $>5$, Item~1c $\ge 66$) are conventional, not derived from data-driven optimisation. Ontological stance was self-reported via a broad binary item and interpreted through a coarse gate for metacognition; nuance is necessarily lost. The instrument’s brevity sacrifices causal richness for transparency and redeployability. Published rubrics risk distortion, especially when indicators acquire institutional salience, a challenge noted in literature \cite{muller2019tyranny}. Mitigations include privileging trends over thresholds, rotating items, tying approval to artefacts rather than ontological stance, and maintaining indirect elicitation.

\subsection{Concluding remarks}

The methodological aim of this study was to prototype a compact indicator set, one that makes policy–practice divergence empirically visible, epistemically legible, and institutionally actionable. The survey itself is illustrative: not a claim about population patterns, but a worked example showing how operational and epistemic signals can be extracted, coded, and interpreted to support assessment pilots, curriculum design, and coherent procurement. The instrument is a reusable scaffold, not a diagnostic device. As generative AI unsettles inherited boundaries between intellect, education, and tools, institutions will need to discern where their policies are actually in effect, and whether academic expertise is being meaningfully applied. The indicators offered here help make that discernment possible.

\section*{Declarations}

\begin{itemize}
  \item \textbf{Ethics approval:} Exempt determination for anonymous, non-interventional survey research.
  \item \textbf{Consent:} Informed by continuation.
  \item \textbf{Competing interests:} None.
  \item \textbf{Data and code:} Anonymised analysis extract and reproducible Python scripts available upon reasonable request.
\end{itemize}

\section*{Funding}
Supported by JSPS Grant-in-Aid for Early-Career Scientists \#24K16628.

\section*{Acknowledgements}
This study owes its clarity to Prof. David Alan Grier and V.E..

\bibliographystyle{apalike}   
\bibliography{main}

\clearpage
\appendix
\section*{Appendix A: Survey items (English wording)}

\begin{enumerate}
\item {For a concrete example of human input and AI output:}
\begin{enumerate}
  \item[\,(1a)\,] Please estimate how many out of 10 complex prompts (inquiries, tasks etc.) that you assigned to AI received satisfactory response on the first try. (0–10)
  \item[\,(1b)\,] Did you need to adapt or adjust the AI’s output (response, solution etc.) that you received before using it? (Always / Frequently / Sometimes / Never)
  \item[\,(1c)\,] If your answer to b) is not "Never", what percentage do you estimate remained AI’s contribution to the result? (0–100)\%
\end{enumerate}

\item Do you see educational value in using AI together with your students during teaching? (Yes / No / I am unsure)

\item If you had the authority to set the policy regarding student use of AI in your course(s), including exams, how would you regulate it? (Fully allow it / Limit its use / Forbid it / I am unsure)

\item Imagine you are having a conversation with someone elsewhere. Do you think that, in a long conversation with twists and turns, you can always distinguish whether the one you are conversing with is human or AI? (Yes / No / I am unsure)

\item Do you think that AI’s capability of supporting the emergence of niche or ultra-specialised fields, through unhindered access to areas of interest and passions not typically promoted through formal education, could challenge the relevance of certain established disciplines? (Yes / No / I am unsure)

\item Has using AI changed the way you engage with your own discipline, such as decreased reliance on traditional means of documentation? (Yes / No / I am unsure)

\item Do you think AI differs from more limited (non-AI) computer technologies, such as search engines, in kind (e.g. AI is a partially humanised computer technology) or only in degree (e.g. AI is just more advanced, not different)? (Only degree / In kind too / I am unsure)

\item Would you be willing to share your work-related AI prompts for research on best AI prompting practices and techniques? (Yes / I need more information / No). If not “No”, please provide your contact information (optional).

\item AI simulates human intellectual performance. Do you think the human intellect is material or immaterial? (Material / Immaterial / I am unsure)

\item In a few sentences, please express your opinion on the relevance of AI for university teachers. (free text)
\end{enumerate}

\end{document}